# High-Magnetic Field Phases in $U_{1-x}Th_xTe_2$

Camilla M. Moir[1], John Singleton[2], Joanna Blawat[2], Eric Lee-Wong[1,3], Yuhang Deng[1], Keke Feng[1], Tyler Wannamaker[1], Ryan E. Baumbach[4,5,§], and M. Brian Maple[1,*]

[1]Department of Physics, University of California, San Diego, La Jolla, CA 92093, USA
[2]National High Magnetic Field Laboratory, Los Alamos National Laboratory, Los Alamos, NM 87545, USA
[3]Department of NanoEngineering, University of California, San Diego, La Jolla, CA 92093, USA
[4]National High Magnetic Field Laboratory, Florida State University, Tallahassee, FL 32310, USA
[5]Department of Physics, Florida State University, Tallahassee, FL 32306, USA

[§]Current address: Department of Physics, University of California, Santa Cruz, Santa Cruz 95064, CA, USA

*Corresponding author: M. Brian Maple. Email: mbmaple@ucsd.edu

**Abstract**

At temperatures much lower than its superconducting critical temperature $T_c$ of 2.1 K, the heavy fermion superconductor $UTe_2$ has a remarkable phase diagram of magnetic field $H$ vs. angles $\varphi$ and $\theta$ at which $H$ is tilted away from the *b*-axis toward the *a*- and *c*-axes, respectively, in the orthorhombic unit cell. The phase diagram appears to contain three distinct superconducting phases: (1) a low field superconducting phase $SC_{LF}$ extending over all values of $\varphi$ and $\theta$ with an upper critical field $H_{c2}$ with a maximum value of 15 T at $\varphi = \theta = 0^o$; (2) a high field superconducting phase $SC_{HF}$ located in a small region between $\varphi \approx 7^o$ and $\theta \approx 4^o$ in fields from $H_{c2LF}$ of the $SC_{LF}$ phase and the metamagnetic transition at $H_m$ at ~35 T marking the onset of the magnetic field polarized FP phase: and (3) a striking $SC_{FP}$ superconducting phase that resides entirely within the FP phase in a pocket of superconductivity extending from $\theta \approx 20^o$ to $40^o$ in fields from ~40 T to above 60 T. In this work, we have studied the evolution of the $H$ vs $\theta$ phase diagram at a base temperature of ~0.6 K as a function of Th concentration $x$ in $U_{1-x}Th_xTe_2$ pseudobinary compounds for 0.5% $\lesssim x \lesssim$ 4.7%. We find that for all values of $x$ within this range, the $SC_{LF}$ phase is retained with a reduced value of $H_{c2}$ of ~10 T at $\varphi = \theta = 0^o$ for $x$ = 4.7%, while the $SC_{HF}$ phase is suppressed. The $SC_{FP}$ and FP phases are unaffected to values of $x$ = 2% but are completely suppressed in the region $x$ = 2.5 to 4.7% where the residual resistance ratio $RRR$ has decreased substantially from ~14 at $x$ = 1.5% to values of ~3, indicating a significant increase in disorder. These results are in marked contrast to the recent studies of non-superconducting disordered $UTe_2$ single crystals in which the $SC_{LF}$ and $SC_{HF}$ phases are absent, but the FP and so-called "orphan" $SC_{FP}$ phases are retained.

## Introduction

There has been an explosion of research on the heavy fermion compound $UTe_2$ following the discovery that it displays an unconventional type of superconductivity [Ran19a, Aoki19a] and multiple superconducting phases induced by application of high magnetic fields [Ran19b, Knebel19] and pressure [Thomas20, Knafo23, Honda23]. At the outset, the unconventional superconductivity was suspected to involve spin-triplet pairing of electrons. This initiated intense experimental and theoretical efforts that are ongoing to develop a fundamental understanding of the unconventional superconductivity of $UTe_2$, [Aoki22, Lewin23]. Interest in $UTe_2$ has also been driven by the possibility that it is a topological superconductor with potential applications for robust quantum computation [Fu08, Sau12].

Since spin-triplet pairing is generally believed to be mediated by ferromagnetic (FM) spin fluctuations [Fay80, Mackenzie03, Kallin09], uranium-based heavy-fermion materials near a FM instability have been considered to be promising candidates for spin-triplet superconductivity. Based on the temperature dependence of its magnetic susceptibility, it was argued that $UTe_2$ is near a FM quantum critical point (QCP) located at the paramagnetic end of a series of FM U-based heavy-fermion superconductors [Ran19a,

Aoki22] which includes $UGe_2$, URhGe, and UCoGe [Aoki19b]. Early experimental evidence suggesting $UTe_2$ is a spin-triplet superconductor includes: (a) Upper critical fields $H_{c2}(T)$ that exceed the Pauli paramagnetic limits along all crystallographic directions [Ran19a, Aoki19a]. (b) STM evidence of chiral-triplet topological superconductivity [Jiao20]; (c) exclusion of spin-singlet pairing from the $^{125}$Te Knight shift reduction below $T_c$ measured by NMR [Nakamine21, Fujibayashi22]; and (d) time reversal symmetry breaking below $T_c$ from a non-zero polar Kerr effect and evidence for two superconducting transitions in the heat capacity [Hayes21, Wei22]. More recent evidence indicates an order parameter symmetry consistent with spin triplet superconductivity from measurements of the Josephson coupling symmetry [Liu24]. On the other hand, recent systematic heat capacity measurements show that $UTe_2$ samples with less disorder have a single superconducting transition at ~2.1 K, a smaller residual resistivity, and a vanishingly small residual Sommerfeld specific heat coefficient in the superconducting state [Rosa22, Aoki22]. Inelastic neutron scattering (INS) studies reveal that superconductivity in $UTe_2$ is coupled to a sharp magnetic excitation, or resonance, at the Brillouin zone (BZ) boundary near AFM order [Duan21, Raymond21]. Since the resonance has only been found in spin-singlet superconductors near an AFM instability, its discovery in $UTe_2$ suggests that AFM spin fluctuations may also induce spin-triplet pairing or that electron pairing in $UTe_2$ has a spin-singlet component. At this point, it seems fair to say that the pairing symmetry of the unconventional superconducting state of $UTe_2$ and the interactions mediating the pairing remain to be established.

At a temperature of ≈0.6 K, well below its zero-field superconducting transition temperature $T_c$ = 2.1 K, $UTe_2$ has an extraordinary phase diagram of magnetic field $H$ vs angles $\varphi$ and $\theta$ at which $H$ is tilted away from the $b$-axis of its body-centered orthorhombic unit cell in the ($b$, $a$) and ($b$, $c$) planes, respectively (note that the $b$- and $a$-axes are the magnetically hard- and easy-axes, respectively) [Ran19b, Knebel19, Aoki22, Lewin23]. This phase diagram contains at least three superconducting phases, which we denote as $SC_{LF}$, $SC_{HF}$, and $SC_{FP}$, and a first-order metamagnetic transition at a magnetic field $H_m$ into a magnetic field polarized phase FP. One of the superconducting phases $SC_{LF}$ has an upper critical field $H_{c2LF}$ ≈ 15 T at $\varphi = \theta = 0^o$, which decreases with increasing $\varphi$ and $\theta$ to values approaching ≈5 T at $\varphi = 90^o$ and ≈10 T at $\theta \approx 90^o$ [Ran19b, Knebel19]. The other two superconducting phases $SC_{HF}$ and $SC_{FP}$ are magnetic field induced phases. The $SC_{HF}$ phase is located near the *b*-axis between the angles $\varphi \approx 7^o$ and $\theta \approx 4^o$ and extends from the $H_{c2LF}(T)$ of the $SC_{LF}$ phase of ≈15 T to $H_m \approx 35$ T which marks the onset of the FP phase. The $SC_{HF}$ phase was initially interpreted as a re-entrant superconducting domain of the $SC_{LF}$ phase [Ran19b, Knebel19], but several subsequent studies of superconductivity in this region suggest the $SC_{HF}$ phase is actually a distinct superconducting phase with an order parameter different than that of the $SC_{LF}$ phase below ≈15 T from which it appears to grow [Knafo21a, Rosuel23, Sakai23, Kinjo23, Lewin24, Wu24]. When the magnetic field is aligned along the *b*-axis, the superconductivity extends to ≈35 T where it is cut off by the first order transition at $H_m$ into the FP phase [Ran19b, Rosuel23, Sakai23, Kinjo23, Wu24]. Even more remarkable is the $SC_{FP}$ phase which forms at $H_m$ and resides within the FP phase in a pocket of superconductivity extending over $\theta$ from ≈20° to ≈45° and $H$ from 40 T to 73 T [Ran19b, Knebel19, Helm24]. It is noteworthy that the $SC_{HF}$ phase for the field along the *b*-axis is quenched at $H_{c2HF}$ by the onset at $H_m$ of the transition into the FP phase, while the $SC_{FP}$ phase for the field oriented at an angle $\theta$ between ≈20° and ≈ 45° has an onset at $H_m$ and resides within the FP phase. Thus, the $SC_{HF}$ phase near the *b*-axis is destroyed by the onset of the FP, while the $SC_{FP}$ phase appears to emerge from the FP phase.

Experiments on $UTe_2$ under high pressure have revealed a $T$ vs $P$ phase diagram consisting of multiple superconducting and magnetic phases. Thomas *et al.* [Thomas20, Thomas21] found that $T_c$ of the ambient pressure superconducting phase decreases linearly with pressure and extrapolates to 0 K near 1.6 GPa, while another superconducting phase emerges from the zero-pressure phase at a pressure of ≈0.2 GPa and has a dome shape with a maximum $T_c$ of ≈3 K at 1.25 GPa and also extrapolates to 0 K near 1.6 GPa. At higher pressure, there is evidence for AFM which appears to emanate from an AFM quantum critical point (QCP) located at a pressure of ~1.3 GPa which lies within the two superconducting phases. Knafo *et al*. [Knafo23] reported similar behavior for the $T_c$ vs $P$ phase boundaries of the two superconducting phases and established that long-range incommensurate antiferromagnetic order occurs at 1.8 GPa by means of neutron-scattering measurements. The antiferromagnetic phase has a propagation vector close to that of the wavevector where antiferromagnetic fluctuations were previously observed at ambient pressure [Duan20, Duan21, Knafo21b, Butch22]. Thus, in $UTe_2$ under pressure, there are two unconventional superconducting phases that apparently envelop an AFM QCP at 1.3 GPa.

X-ray diffraction (XRD) measurements in diamond anvil cells (DACs) at high pressures reveal that $UTe_2$ undergoes an orthorhombic to tetragonal structural phase transition between 5 and 7 GPa, with a large volume collapse of nearly 10 % and an increase in nearest U-U distance by about 4 %. [Aoki22, Huston22, Honda23, Deng24]. No new structural transitions were found beyond 7 GPa up to 30 GPa [Deng24]. This lower to higher symmetry transition suggests less 5*f* electron participation in bonding in the tetragonal phase of $UTe_2$. The scenario of increased localization of 5*f*-electrons in $UTe_2$ under pressure is supported by measurements of the electrical resistivity as a function of temperature reported by Honda *et al*. [Honda23]. These measurements reveal the occurrence of superconductivity above 6 GPa with an upper critical field lower than the Pauli limit and Fermi-liquid behavior in the normal state electrical resistivity $\rho(T)$ with a small coefficient ($A$) of the $T^2$ term, indicating that the electronic state of tetragonal $UTe_2$ is weakly correlated [Honda23]. This is consistent with the behavior of $\rho(T)$ which has a shoulder at ≈230 K, suggesting a coherence temperature $T^* \approx 230$ K in the tetragonal phase [Honda23], which is significantly larger than the value of $T^*$ (10-70 K) in the orthorhombic phase [Ran19a, Eo22]. The rich variety of superconducting and magnetic phases that are induced by magnetic field and pressure in $UTe_2$ provides an opportunity to explore the phase space defined by temperature, magnitude and orientation of magnetic field, and pressure. This interesting direction has been pursued in several recent investigations [Ran21a, Aoki22, Knebel23].

The effect of atomic disorder on the superconducting phases of $UTe_2$ in zero and high magnetic fields has been studied in recent experiments. The zero-field $T_c$ of $UTe_2$ has been reported to be very sensitive to tellurium deficiency [Cairns20] which can be controlled by adjusting the temperature gradient used in chemical vapor transport (CVT) crystal growth or by changing the growth method [Cairns20, Ran21b, Rosa22, Sakai22, Aoki24]. Additionally, recent experiments investigating CVT and molten-salt grown $UTe_2$ have found that a small U deficiency of ≈ 4 - 5 % is enough to completely suppress the low-field superconducting phase [Haga22, Rosa22]. Interestingly, experiments by Frank *et al.* [Frank23] on samples with very large disorder in high magnetic fields found the striking result that disorder suppresses the $SC_{LF}$ and $SC_{HF}$ phases, but preserves the FP and $SC_{FP}$ phases, implying that the $SC_{LF}$, $SC_{HF}$ and $SC_{FP}$ phases may arise from different mechanisms or even be in competition with one another [Frank23]. This so-called $SC_{FP}$ "orphan superconductivity" exists at angles $\theta$ between 29° and 42° for applied magnetic fields between 37 T and 52 T. The non-superconducting $UTe_2$ crystals that exhibit the $SC_{FP}$ "orphan superconductivity" have a high degree of disorder as evidenced by the residual resistance ratio of RRR $\approx$ 7. The largest value of RRR that has been reported is ≈1000 for $UTe_2$ prepared by growth in a molten salt flux [Sakai22, Wu24].

The objective of the research reported herein was to study the effect of Th substitution for U in $UTe_2$ on the high field phases $SC_{LF}$, $SC_{HF}$, $SC_{FP}$ and FP as function of Th substituent concentration, temperature, magnetic field $H$, and angle $\theta$ that $H$ is tilted away from the $b$-axis in the ($b$, $c$) plane. The proximity diode oscillator (PDO) technique was used to make contactless conductance measurements on $U_{1-x}Th_xTe_2$ single crystals with Th concentrations up to $x \approx 0.47$ over the temperature range ≈0.6 K (base temperature) to ≈2.1 K in magnetic fields up to 60 T at the National High Magnetic Field Laboratory (NHMFL) at Los Alamos National Laboratory (LANL). The substitution of Th is expected to weaken the electronic correlations in $UTe_2$ by introducing disorder, decreasing the number of $f$-electrons, decreasing the number of U dimers, and increasing the itinerant electron concentration. Rosa *et al.* [Rosa22] found that substitution of Th for U in $UTe_2$ produces a strong depression of the zero field $T_c$ of $UTe_2$ that appears to vanish by $x$ = 10% Th. For low Th content ($x$ = 0.5% to 2%), we observed the $SC_{LF}$ phase below 10 T, the metamagnetic transition $H_m$ to the FP phase, as well as the $SC_{FP}$ phase at roughly 35° from the $b$-axis. However, we did not observe the $SC_{HF}$ phase that has been reported for the parent compound $UTe_2$ for $\theta \lesssim 4°$. For higher thorium content ($x$ = 2.5% and 4.7%), we did not observe the FP, $SC_{FP}$, or $SC_{HF}$ phases and were only able to detect the $SC_{LF}$ phase. This behavior is in marked contrast to that reported in the study of non-superconducting disordered $UTe_2$ single crystals by Frank *et al.* [Frank23] discussed above, in which it was found that the $SC_{LF}$ and $SC_{HF}$ phases had been suppressed, but the "orphan" FP and $SC_{FP}$ phases had been preserved. We will discuss the results in terms of the possible role of U dimers, the exchange field compensation effect (Jaccarino-Peter effect), and spin-triplet superconductivity in the discussion section of this paper.

**Experimental Methods**

Single crystals of $U_{1-x}Th_xTe_2$ were grown by chemical vapor transport. Tellurium (99.99% metal basis), thorium-232, and depleted uranium were added to a quartz tube in a 2:3 atomic ratio of U/Th:Te. Samples ranged in nominal thorium content from $x \approx 0.005$ to 0.047. Iodine was added in an amount equivalent to 4 mg/cm$^3$ relative to the volume of the quartz tube, and all elements were sealed in an argon environment with a gas pressure below 1 Torr. Sealed ampoules were fired under temperature gradients of 800 °C to 700 °C for $x$ = 0.005, 0.0096, 0.015, 0.020, 0.025, 0.027, 0.030, 0.040, 0.047, and 1060 °C to 1000 °C for $x$ = 0.0129, for two weeks. The resulting crystals ranged in size from 1 mm to 10 mm.

Sample quality was assessed by comparison of Laue images with those expected for the structure of $UTe_2$, sharpness of the "specific heat jump" at $T_c$, and energy dispersive spectroscopy (EDS). Laue images were taken using a Photonic Science Laue Orientation System at ambient pressure and temperature. Specific heat $C(T)$ measurements were performed in a Quantum Design DynaCool Physical Property Measurement System (PPMS) at temperatures down to 1.8 K. The $C(T)$ data displayed in Figure 1 show that $T_c$ decreases with $x$, except for the sample with $x$ = 0.0129 which was grown at a high temperature gradient that is known to produce $UTe_2$ samples with lower values of $T_c$ [Rosa22]. The $T_c$ of the $x$ = 0.0129 sample is below the temperature range of the DynaCool PPMS and was determined by PDO measurements at the LANL NHMFL (see below).

The EDS measurements were performed with an FEI scanning electron microscope on the $U_{1-x}Th_xTe_2$ crystal with the largest nominal Th concentration $x$ = 0.047 to determine the relation between the nominal and actual thorium content. Analysis of EDS spectra taken at three points on the crystal yielded an average value of the thorium concentration within 10% of the nominal concentration. Since the analysis of the specific heat measurements, described below, revealed that the values of $T_c$ of the samples studied in this work scaled with the nominal Th concentration, the nominal and actual Th concentrations were taken to be equal based on the small discrepancy between the nominal and actual Th content for the *x* = 0.047 crystal.

All of the measurements on the $U_{1-x}Th_xTe_2$ single crystals in high magnetic fields reported in this paper were performed by means of the PDO contactless conductivity technique at the NHMFL at LANL. The circuit used for PDO measurements is described in references [Singleton00, Altarawneh09, Ghannadzadeh11].

Electrical resistivity $\rho(T)$ measurements were made between 300 K and 1.8 K in the DynaCool PPMS on selected samples with the current flowing along the *b*-axis. The $\rho(T)$ curves for selected values of x closely resemble the $\rho(T)$ curve for $UTe_2$ reported by Ran *et al.* [Ran19a] which exhibits a maximum near 75 K below which there is a rapid drop in $\rho(T)$ due to the formation of the heavy fermion ground state, culminating in an abrupt drop in $\rho(T)$ to zero at $T_c$ due to the onset of superconductivity. The residual resistance ratio *RRR* defined as *R*(300 K)/*R*(2.5 K), where *R* is the electrical resistance, decreases with *x* and has the following values: *RRR* = 24.5 (*x* = 0), 21.3 (*x* = 0.005), 14.3 (*x* = 0.015), 3.2 (*x* = 0.025), 2.5 (*x* = 0.047).

**Experimental Results**

Figure 1(a) shows the normalized heat capacity divided by temperature (*C*/*T*) as a function of temperature (*T*) for $U_{1-x}Th_xTe_2$ at various nominal Th concentrations *x* ranging from 0 to 0.047. For the $UTe_2$ parent compound, there is a prominent “heat capacity jump” in *C*/*T* at $T_c$ = 1.99 K, where $T_c$ is defined as the intercept on the *T*-axis of a straight line that is tangent to the *C*/*T* curve in the region of the “jump” in C/*T*, which is near the onset of the superconducting transition. This definition of $T_c$ was adopted in order to determine the shift in $T_c$ with x from the jumps in the *C*(*T*)/*T* curves in Figure 1, some of which were cut off at low temperature by the 1.8 K low temperature limit of the DynaCool PPMS. For the $U_{1-x}Th_xTe_2$ single crystals, the jump in *C*/*T* and, in turn, $T_c$, shifts to lower temperature with increasing *x*, as expected from previous measurements by Rosa *et al.* [Rosa22]. The slope of the jump in *C*/T and, in turn, the width of superconducting transition $\Delta T_c$ is nearly constant in this Th concentration *x* range. Shown in Figure 1(b) is a plot of $T_c$ vs. Th concentration *x* in which a linear fit to the data yields a rate of depression $dT_c/dx \approx$ -0.03 K/atm.% Th. This is about an order of magnitude smaller than the value reported in ref. [Rosa22], which may be due to differences in the gradient end point temperatures used in the CVT crystal growth process. No data for the crystal with $x = 0.0129$ are shown in Figure 1(a) or 1(b) since, as noted above, it had a $T_c$ of 1.6 K which is below the low temperature limit of the DynaCool PPMS (1.8 K).

Figure 2 shows a PDO measurement frequency curve as a function of magnetic field for a $U_{1-x}Th_xTe_2$ single crystal with $x_{nom} = 0.02$. The phase transitions, marked by a sudden change in frequency, are indicated by arrows. Since the PDO frequency shift is inversely proportional to the changes in the resistance, an increase in the frequency indicates a drop in resistance or the onset of superconductivity. A decrease in the frequency can indicate either the suppression of superconductivity or the onset of the metamagnetic transition, depending on the field strength. In Figure 2, the data taken at 34° from the *b*-axis clearly show three transitions: a suppression of the $SC_{LF}$ phase at 8.26 T, the development of the $SC_{FP}$ phase at 42.3 T, and the suppression of the $SC_{FP}$ phase at 52.3 T. In contrast, at 20°, we are outside the range of angles where the $SC_{FP}$ phase exists and we only observe the suppression of the $SC_{LF}$ phase at $H_{c2LF} = 9.9$ T and the metamagnetic transition field $H_m$, the onset of the FP phase, at 35.5 T.

The evolution of the phase diagrams of $H$ vs. angle $\theta$ of $H$ from the *b*-axis in the (*b*, *c*) plane as thorium content increases is summarized in nine phase diagrams from $x = 0$ to 0.047 in Figure 3 (a-i). The high field phases are clearly visible for the lower thorium concentrations ($x$ = 0.005, 0.0096, 0.0129, and 0.02), with the exception of the $SC_{HF}$ phase that occurs close to the *b*-axis in the parent compound. For $x = 0.005$, very close to the *b*-axis, we see an increase in the frequency, similar to a signal indicating the onset of superconductivity at magnetic fields above which we observe the decrease in frequency associated with $H_{c2HF}$ (Fig 4 (a)). It is possible that this is a remnant of the $SC_{HF}$ phase that occurs close to the *b*-axis in the parent compound $UTe_2$. If this is the case, then the $SC_{HF}$ *b*-axis phases are the most susceptible to thorium substitution, and it is possible that very small levels of thorium may suppress the *b*-axis $SC_{HF}$ phase. It is important to note, however, that at $\theta = 7.5$, the frequency change possibly due to a remnant $SC_{HF}$ phase occurs at a higher field than the frequency decrease due to the onset of the field polarized phase and warrants further investigation. The FP phase and $SC_{FP}$ phase for $x = 0.0129$ are compressed with respect to the *b*-axis (Fig 3 (c)) which can be contrasted with the high field phases for $x = 0.0096$ (Fig 3 (b)) and $x = 0.02$ (Fig 3 (d)) which have broader FP phases with angular dependences much closer to that of the parent compound [Ran19b]. The contraction of the high field phases with respect to the *b*-axis is most likely due to a misalignment of the rotation angle to the crystallographic angle, such that $\theta$ describes a line from the *b*-axis to some point between the *c*- and *a*-axes. Previous work by Ran *et al*. on the parent compound shows the FP phase from the *b*-axis to the *a*-axis is not as wide in angle as the phase between the *b*- and *c*-axes [Ran19b]. However, we also note that the sample with $x = 0.0129$ was grown with a higher temperature gradient than the other samples in the substitution series. It is possible that differences in the growth conditions that produce crystals with lower values of zero field $T_c$ [Rosa22] are also responsible for the unexpected contraction of the high field phases as a function of $\theta$; however, the $H$ vs $\theta$ phase diagrams of disordered $UTe_2$ reported by Frank *et al.* do not exhibit the contraction towards the *b*-axis observed in the $x = 0.0129$ crystal [Frank23].

In addition to the observed high field phases, starting with $x = 0.02$ (Fig 3(d)), we notice a local minimum in $H_{c2LF}$ with $\theta$. As with the parent compound and all samples with $x \leq 0.02$, $H_{c2HF}$ is at a maximum when the field is along the *b*-axis [Ran19b, Rosuel23, Sakai23, Kinjo23, Wu24]. As $\theta$ is increased, we see a local minimum near 40° that coincides with the maximum in $H_{c2FP}$ (gray dashed line). If the $SC_{FP}$ and $SC_{LF}$ phases are in competition with each other, it naïvely makes sense that the $SC_{LF}$ phase would be the weakest at the angle where the $SC_{FP}$ phase was the strongest. As the thorium content increases, the local minimum becomes broader, such that the $SC_{LF}$ phase separates into two distinct lobes that are centered on the *b*- and *c*-axes.

We note that at $x = 0.025$, we do not observe a clear metamagnetic transition; however, we do detect a broad feature that appears as an 'elbow' in the frequency as a function of magnetic field (Fig 4 (b)). If we track this broad feature, it appears to trace out a similar area in the $H$ vs $\theta$ plot as the FP phase. While it is not clear what this feature represents, it is possible that the system is near the critical value of $x$ where $H_m$ vanishes and has only partially entered the FP phase. If this is true, then the $SC_{FP}$ phase, which is not observed for $x = 0.025$, is suppressed before the metamagnetic transition with increasing

Th substitution. Both the FP and $SC_{FP}$ phases are absent for values of $x$ between 0.025 and 0.047.

The temperature dependence at a fixed angle of 33° in the range of the $SC_{FP}$ phase is shown in Figure 5 for $x$ = 0.005, 0.0096, 0.0129, and 0.02. For all Th substituent concentrations $x$, the angle of 33° at which the temperature dependence was measured was selected so that the isothermal magnetic field sweeps would intersect the $SC_{LF}$, FP, and $SC_{FP}$ phases. As the temperature increases, $H_{c2LF}$ and $H_{c2FP}$ decrease in magnetic field for all substitutions. The onset field of the $SC_{FP}$ phase does not change with temperature as it is pinned to the metamagnetic transition at $H_m$ which has no obvious temperature dependence within the temperature range measured (roughly 2 K). We do note that for all substitutions measured, the $SC_{FP}$ phase outlives the $SC_{LF}$ phase as the temperature is increased (see $T$ = 1.83 K, 1.93 K, 1.62 K, and 1.88 K for $x$ = 0.005, 0.0096, 0.0129, and 0.02, respectively).

**Discussion**

Electronic correlations are expected to be weakened in Th substituted $UTe_2$ by introducing disorder, decreasing the number of $f$-electrons, decreasing the number of U dimers [Christovam24], and increasing the itinerant electron concentration, consistent with the depression of the zero field $T_c$ of $UTe_2$ with Th substituent concentration. It is noteworthy that the $T_c$ of $UTe_2$ decreases with U deficiency [Haga22]. Our studies of the evolution of the low temperature $H$ vs $\theta$ phase diagram of $U_{1-x}Th_xTe_2$ in the range 0.5% $\lesssim x \lesssim$ 4.7% reveal that Th substitution also affects the magnetic field dependence of the superconducting $SC_{LF}$, $SC_{HF}$ and $SC_{FP}$ phases, as well as the FP phase. Within this Th composition range, the general behavior of the $SC_{LF}$ phase is maintained throughout the range 0° $\lesssim \theta \lesssim$ 90°, but with a reduced value of $H_{c2LF}$ at $\theta$ = 0 and 0.6 K (e.g., $H_{c2LF} \approx$ 15 T at $x$ = 0 and $\approx$ 5 T at $x$ = 4.7%). In contrast, the $SC_{HF}$ phase for $H$ aligned in the narrow range of angles $\theta \lesssim$ 4° is very sensitive to Th substitution and is completely suppressed at $x \geqslant$ 0.5%.

The much greater sensitivity to Th substitution of the $SC_{HF}$ phase compared to the $SC_{LF}$ phases provides additional evidence that these phases are two distinct superconducting phases, with different pairing symmetries and/or mechanisms, in support of several detailed studies of the superconducting properties of $UTe_2$ in high magnetic fields for *H* aligned near the *b*-axis [Knafo21a, Rosuel23, Sakai23, Kinjo23, Lewin24, Wu24]. The $SC_{FP}$ superconducting pocket and the FP phase are only weakly affected by Th substitution up to $x = 0.02$ but are completely suppressed for Th concentrations $x = 2.5$ to 4.7% where the *RRR* has decreased substantially from ~14 at $x = 1.5\%$ to values of ≈3 in this range, indicating a significant increase in disorder. These results are in marked contrast to the recent studies of non-superconducting disordered $UTe_2$ single crystals in which the $SC_{LF}$ and $SC_{HF}$ phases are absent [Cairns20, Ran21b, Haga22, Rosa22, Sakai22, Frank23, Aoki24], but the FP and so-called "orphan" $SC_{FP}$ phases are retained [Frank23]. In summary, substitution of Th into $UTe_2$ up to 4.7% first suppresses the $SC_{HF}$ phase and then the FP and $SC_{FP}$ phases, and finally reduces $H_{c2LF}$ of the $SC_{LF}$ phase; sufficient U or Te vacancy disorder in $UTe_2$ completely destroys the $SC_{LF}$ and $SC_{HF}$ phases [Cairns20, Ran21b, Haga22, Rosa22, Sakai22, Frank23, Aoki24] but preserves the FP and "orphan" $SC_{FP}$ phases.

*Uranium Dimers*

The structure of $UTe_2$ contains a network of two-leg ladders extending in the *a*-direction in which pairs of U atoms, or U dimers with the shortest U-U distance, form the rungs of the ladders oriented in the *c*-direction. Inelastic neutron scattering measurements performed in the normal state of $UTe_2$ above $T_c$ have provided evidence for FM exchange interactions between the U ions in the dimers that form the rungs of the ladder and AFM exchange interactions between U atoms in neighboring ladders [Knafo21b]. It is clear that the substitution of Th for U in $UTe_2$ disrupts and suppresses the high magnetic field $SC_{HF}$, FP, and $SC_{FP}$ phases to a far greater extent than the low field $SC_{LF}$ phase. If we assume that the U dimers and their collective magnetic moments are involved in the formation of these high field phases, then we can also draw some limited conclusions about the dependence of these phases on the dimerization. The higher field $SC_{HF}$ phase for *H* aligned near the *b*-axis is very sensitive to Th substitution, indicating that it is highly dependent on U dimers or very sensitive to disorder in the system. On the other hand, there is no perceptible change in the ≈35 T metamagnetic transition to the FP phase with Th substitution until it reaches a critical concentration above which the metamagnetic transition, the FP phase, and the $SC_{FP}$ phase completely disappear. This may be due to some critical limit of available U dimers or a possible change in the crystal structure. However, the Th content even at $x = 0.04$ is very small and no changes in $UTe_2$ patterns were seen in the Laue images making the structural change scenario unlikely (see Fig. 6). It is noteworthy that FM intra-unit cell interactions between U ions within a U dimer have been considered as possible mechanisms for spin triplet pairing in $UTe_2$ [Chen21, Shishidou21]. Chen *et al.* [Chen21] have proposed that AFM inter-unit cell exchange interactions could account for the incommensurate AFM fluctuations observed in INS experiments [Duan20, Duan21, Knafo21b, Butch22] and the resonance in the spin spectrum of $UTe_2$ [Duan21, Raymond21], as well as the incommensurate AFM ordering under pressure [Knafo23]. If FM-ordered U dimers are responsible for the spin-triplet $SC_{LF}$, based on the fact that $SC_{LF}$ is not very sensitive to Th, Th apparently does not have much of an affect the FM-U dimers.

*Jaccarino-Peter (J-P) Effect*

Another possibility is that the $SC_{FP}$ phase which emerges from the FP phase involves the same kind of superconductivity as the $SC_{LF}$ phase which reappears at high fields in the form of the $SC_{FP}$ phase due to the interplay of the magnetic field $H$ and interactions associated with the FP phase. One such possibility is the exchange field compensation effect, also known as the Jaccarino-Peter effect [Jaccarino62]. The usual formulation of the J-P effect is based on the compensation of the applied magnetic field by a negative exchange field associated with a subset of ions that carry magnetic moments in a superconducting material in which the paramagnetic limiting field $H_P(T)$ is smaller than the orbital critical field $H^*(T)$. Coupling between these magnetic moments and the conduction electron spins generates an “effective magnetic field,” or exchange field $H_J$, that acts on the conduction electron spins in the same manner as an applied magnetic field. If the sign of the coupling between the localized magnetic moments and the conduction electron spins is negative, the direction of $H$ will be opposite to that of $H_J$; that is, the effect of $H_J$ will be “compensated” by $H$. The net magnetic field $H_T$ is then given by $H_T = H - |H_J|$ and the material will be superconducting (S) when $|H_T| < H_P$ and normal (N) when $|H_T| > H_P$. The exchange field $H_J$ is proportional to the Brillouin function $B_J(g_J J \mu_B H / k_B T)$ for paramagnetic localized moments with total angular momentum J and Landé g-factor $g_J$, which increases in magnitude with increasing *H* and decreasing *T*. Thus, the *H* and *T* dependence of $H_J$ affect the temperature dependence of the upper critical field curve $H_{c2}(T)$ so that it develops regions with positive curvature or even becomes re-entrant wherein at certain values of the magnetic field there are two superconducting domains, one at lower fields (starting from 0 field) and another at higher fields. Thus, at the lowest temperature, there is a succession of transitions with increasing magnetic field – from superconducting to normal, to superconducting again, and finally, back to normal (S-N-S-N). A simple pictorial explanation of the J-P effect can be found in references [Maple85] and [Wolowiec15]. The J-P effect has been used to describe the $H_{c2}(T)$ curves of various Chevrel phase superconductors [Maple85] with regions of positive curvature such as $EuMo_6S_8$ [Decroux84], re-entrant superconductivity with two superconducting domains such as $Eu_{0.75}Sn_{0.25}Mo_6S_{7.2}Se_{0.8}$ [Meul84], and magnetic field induced superconductivity in the antiferromagnetic insulator λ-$(BETS)_2FeCl_4$ [Uji01]. In the latter case, the low field superconducting phase is not realized because it is suppressed by other competing interactions that render it an insulating antiferromagnet. In fact, Jaccarino and Peter originally suggested that the exchange field compensation effect could lead to magnetic-field-induced superconductivity (MFIS) in a weakly ferromagnetic material, assuming that it would be superconducting in the absence of ferromagnetic ordering. However, as noted in reference [Maple85], MFIS in a ferromagnet remains to be discovered.

This J-P effect has recently been considered by Helm *et al.* [Helm24] as a possible explanation for the appearance of the $SC_{FP}$ pocket of superconductivity for $\theta$ between ≈25° and ≈45° where the exchange field in the case of $UTe_2$ would be associated with the FP phase. Helm *et al*. found evidence supporting this interpretation in a strong suppression of the normal-state Hall effect that is indicative of a reduced band polarization above $H_m$ in the angular range around 30° due to partial compensation of the applied field by an exchange field. In this scenario, the $SC_{FP}$ phase would involve the type of superconductivity exhibited by the $SC_{LF}$ phase which would reappear at higher fields due to the reduction of the net field acting on the spins of conduction electrons by a negative exchange field associated with the FP phase. Also, since the $SC_{LF}$ phase in the Te deficient $UTe_2$ crystals studied by Frank *et al.* [Frank23] has been suppressed, presumably by disorder, there is no superconductivity to restore at higher fields in the $SC_{FP}$ phase by the exchange field compensation effect associated with the FP phase, unless the basic interactions responsible for superconductivity in the $SC_{LF}$ phase could somehow be activated at a field below $H_m$. However, it should be noted that $UTe_2$ may still be proven to be a spin-triplet superconductor [Aoki23, Liu24] and the exchange field compensation effect in its original form relies on $UTe_2$ being a spin-singlet superconductor.

Interestingly, re-entrant superconductivity has also been observed as a function or temperature in a fixed magnetic field in magnetic superconductors containing lanthanide (*Ln*) ions such as $NdRh_4B_4$ due to features in the critical magnetic field $H_{c2}(T)$ associated with the onset of antiferromagnetic ordering of the *Ln* ions that coexist with superconductivity [Hamaker79, Wolowiec15].

*Competition between $SC_{FP}$ and $SC_{LF}$ phases*

The local minima of $H_{c2LF}$ close to the maxima of $H_{c2FP}$ is interesting and suggests the possibility that the two superconducting phases are in competition. However, the suppression of $H_{c2LF}$ between the *b*- and *c*-axes continues after the $SC_{FP}$ and FP phases are no longer observable. It is possible that the $SC_{FP}$ and FP phases have shifted upward in magnetic field, but are still positioned in the same location with respect to angle. In this case, we would be unable to explain the broad feature observed in the $U_{1-x}Th_xTe_2$ single crystal with $x_{nom}$ = 0.025. The other possibility is that the interactions which produce the $SC_{LF}$ phase are still present and depress the $T_c$ of the zero-magnetic field superconducting phase, but are not able to generate the high magnetic field $SC_{FP}$ superconducting phase.

*Spin-triplet pairing*

The most intriguing possibility is that the $SC_{FP}$ phase which emerges from the FP phase is a spin-triplet phase, the origin of which depends on interactions associated with the FP phase. As discussed above, this could be due to a FM exchange interaction within the U-dimers, which is interrupted by Th substitution above a certain concentration. This would also be consistent with the experiments of Frank *et al.* [Frank23] in which disorder was found to destroy the $SC_{LF}$ and $SC_{HF}$ phases and preserve the $SC_{FP}$ and FP phases.

**Concluding remarks**

Our studies of the evolution of the low temperature *H* vs $\theta$ phase diagram of $U_{1-x}Th_xTe_2$ in the range 0.5% $\leqslant$ $x$ $\leqslant$ 4.7% reveal that Th substitution affects the magnetic field dependence of the superconducting $SC_{LF}$, $SC_{HF}$ and $SC_{FP}$ phases, as well as the FP phase. The $SC_{LF}$ phase is preserved with a decreased value of $H_{c2LF}(T)$ for all values of $\theta$ within this Th composition range, the $SC_{HF}$ phase for *H* oriented near the *b*-axis is suppressed for $x \gtrsim$ 0.5%, and the FP and $SC_{FP}$ phase are preserved for 0.5% ≤ $x$ ≤ 2% and suppressed in samples with 2.5% $\leqslant$ $x$ $\leqslant$ 4.7%. These results suggest that the $SC_{LF}$ and $SC_{HF}$ phases are distinct superconducting phases with different types of pairing symmetry/mechanisms, in accord with other detailed experiments discussed above. The $SC_{FP}$ phase that emerges from the FP phase in a superconducting pocket for 20° $\lesssim \theta \lesssim$ 40° in fields above 40 tesla could arise in two ways: (1) It could be produced by electron pairing interactions associated with the FP phase or (2) be restored from the $SC_{LF}$ phase by the interplay of the FP phase and the magnetic field *H* such as the exchange field compensation effect discussed above. Scenario (1) would seem to favor spin-triplet superconductivity and could involve the U-dimers as discussed above. Interruption of the U-dimers by substitution of Th for U could lead to the suppression of the FP and $SC_{FP}$ phases at some critical value of Th concentration as occurs above x = 2%. Scenario (2) would seem to favor spin-singlet superconductivity if it were to be associated with the exchange field compensation effect which implies that the $SC_{LF}$ phase involves spin-singlet superconductivity. Interestingly, this would be at variance with the suggestion by Rosuel *et al*. [Rosuel23] that the $SC_{LF}$ phase is associated with spin-triplet pairing. Thus, it is not clear whether the $SC_{LF}$ and $SC_{FP}$ phases are related to one another. Our results on Th-substituted $UTe_2$ single crystals are in marked contrast to those found in recent studies of non-superconducting disordered $UTe_2$ single crystals in which the $SC_{LF}$ and $SC_{HF}$ phases are suppressed, but the FP and so-called “orphan” $SC_{FP}$ phases are preserved. The experiments reported in this paper will hopefully be useful in elucidating the relationship between the superconducting and FP phases in a magnetic field and the electron pairing symmetry and mechanism responsible for the extraordinary superconducting properties of $UTe_2$.

## Acknowledgments

Research at the University of California, San Diego was supported by the National Nuclear Security Administration (NNSA) under the Stewardship Science Academic Alliance Program through the US DOE under Grant DE-NA0004086, and the US Department of Energy (DOE) Basic Energy Sciences under Grant DE-FG02-04ER46105. This work was sponsored in part by the UC San Diego Materials Research Science and Engineering Center (UCSD MRSEC), supported by the National Science Foundation (NSF) under Grant DMR-2011924. Research at Florida State University and the National High Magnetic Field Laboratory was supported by NSF Cooperative Agreement DMR-2128556, and the DOE. J. S. thanks the Department of Energy Basic Energy Sciences Field Work Proposal *Science of 100 T* for support. This work was performed in part at the San Diego Nanotechnology Infrastructure (SDNI) of UCSD, a member of the National Nanotechnology Coordinated Infrastructure, which is supported by the National Science Foundation (Grant ECCS-2025752).

**References (**High-Magnetic Field Phases in $U_{1-x}Th_xTe_2$, C. M. Moir *et al.*)

[Altarawneh09] M. M. Altarawneh, C. H. Mielke, and J. S. Brooks, "Proximity detector circuits: an alternative to tunnel diode oscillators for contactless measurements in pulsed magnetic field environments," *Rev. Sci. Instrum.* **80**, 066104 (2009).

[Aoki19a] Dai Aoki, Ai Nakamura, Fuminori Honda, DeXin Li, Yoshiya Homma, Yusei Shimizu, Yoshiki J. Sato, Georg Knebel, Jean-Pascal Brison, Alexandre Pourret, Daniel Braithwaite, Gerard Lapertot, Qun Niu, Michal Vališka, Hisatomo Harima, and Jacques Flouquet, "Unconventional superconductivity in heavy fermion $UTe_2$," *J. Phys. Soc. Jpn.* **88**, 043702 (2019).

[Aoki19b] D. Aoki, K. Ishida and J. Flouquet, "Review of U-based ferromagnetic superconductors: comparison between $UGe_2$, URhGe, and UCoGe," *J. Phys. Soc. Jpn.* **88**, 022001 (2019).

[Aoki22] D. Aoki, J-P. Brison, J. Flouquet, K. Ishida, G. Knebel, Y. Tokunaga, and Y. Yanase, "Unconventional superconductivity in $UTe_2$," *J. Phys.: Condens. Matter* **34**, 243002 (2022).

[Aoki24] Dai Aoki, "Molten Salt Flux Liquid Transport Method for Ultra Clean Single Crystals $UTe_2$," *J. Phys. Soc. Jpn.* **93**, 043703 (2024).

[Butch22] Nicholas P. Butch, Sheng Ran, Shanta R. Saha, Paul M. Neves, Mark P. Zic, Johnpierre Paglione, Sergiy Gladchenko, Qiang Ye, and Jose A. Rodriguez-Rivera, "Symmetry of magnetic correlations in spin-triplet superconductor $UTe_2$," *npj Quantum Mater.* **7**, 39 (2022).

[Cairns] Luke Pritchard Cairns, Callum R. Stevens, Christopher D. O'Neill and Andrew Huxley, "Composition dependence of the superconducting properties of $UTe_2$," *J. Phys.: Condens. Matter* **32**, 415602 (2020).

[Chen21] Lei Chen, Haoyu Hu, Christopher Lane, Emilian M. Nica, Jian-Xin Zhu, and Qimiao Si, "Multiorbital spin-triplet pairing and spin resonance in the heavy-fermion superconductor $UTe_2$," *arXiv preprint arXiv:2112.14750* (2021).

[Christovam24] Denise S. Christovam, Martin Sundermann, Andrea Marino, Daisuke Takegami, Johannes Falke, Paulius Dolmantas, Manuel Harder, Hlynur Gretarsson, Bernhard Keimer, Andrei Gloskovskii, Maurits W. Haverkort, Ilya Elfimov, Gertrud Zwicknagl, Alexander V. Andreev, Ladislav Havela, Mitchell M. Bordelon, Eric D. Bauer, Priscila F. S. Rosa, Andrea Severing, and Liu Hao Tjeng, "Stabilization of U $5f^2$ configuration in $UTe_2$ through U 6d dimers in the presence of $Te_2$ chains," *Phys. Rev. Res.* **6**, 033299 (2024).

[Decroux84] M. Decroux, S. E. Lambert, M. S. Torikachvili, M. B. Maple, R. P. Guertin, L. D. Woolf, and R. Baillif, “Observation of Bulk Superconductivity in $EuMo_6S_8$ under Pressure,” *Phys. Rev. Lett.* **52**, 1563 (1984).

[Deng24] Yuhang Deng, Eric Lee-Wong, Camilla M. Moir, Ravhi S. Kumar, Nathan Swedan, Changyong Park, Dmitry Yu Popov, Yuming Xiao, Paul Chow, Ryan E. Baumbach, Russell J. Hemley, Peter S. Riseborough, and M. Brian Maple, “Structural transition and uranium valence change in $UTe_2$ at high pressure revealed by x-ray diffraction and spectroscopy,” *Phys. Rev. B* **110**, 075140 (2024).

[Duan21] Chunruo Duan, R. E. Baumbach, Andrey Podlesnyak, Yuhang Deng, Camilla Moir, Alexander J. Breindel, M. Brian Maple, E. M. Nica, Qimiao Si, and Pengcheng Dai, “Resonance from antiferromagnetic spin fluctuations for superconductivity in $UTe_2$,” *Nature* **600**, 636 (2021).

[Duan20] Chunruo Duan, Kalyan Sasmal, M. Brian Maple, Andrey Podlesnyak, Jian-Xin Zhu, Qimiao Si, and Pengcheng Dai, “Incommensurate Spin Fluctuations in the Spin-triplet Superconductor Candidate $UTe_2$,” *Phys. Rev. Lett.* **125**, 237003 (2020).

[Eo22] Y. S. Eo, S. Liu, S. R. Saha, H. Kim, S. Ran, J. A. Horn, H. Hodovanets, J. Collini, T. Metz, W. T. Fuhrman, A. H. Nevidomskyy, J. D. Denlinger, N. P. Butch, M. S. Fuhrer, L. A. Wray, and J. Paglione, “*c*-axis transport in $UTe_2$: Evidence of three-dimensional conductivity component,” *Phys. Rev. B* **106**, L060505 (2022).

[Fay80] D. Fay and J. Appel, “Coexistence of p-state superconductivity and itinerant ferromagnetism,” *Phys. Rev. B* **22**, 3173 (1980).

[Frank23] Corey E. Frank, Sylvia K. Lewin, Gicela Saucedo Salas, Peter Czajka, Ian M. Hayes, Hyeok Yoon, Tristin Metz, Johnpierre Paglione, John Singleton, and Nicholas P. Butch, “Orphan high field superconductivity in non-superconducting uranium ditelluride,” *Nat. Commun.* **15**, 3378 (2024).

[Fu08] L. Fu and C. L. Kane, “Superconducting proximity effect and Majorana fermions at the surface of a topological insulator,” *Phys. Rev. Lett.* **100**, 096407 (2008).

[Fujibayashi22] H. Fujibayashi, G. Nakamine, K. Kinjo, S. Kitagawa, K. Ishida, Y. Tokunaga, H. Sakai, S. Kambe, A. Nakamura, Y. Shimizu, Y. Homma, D. Li, F. Honda, and D. Aoki, “Superconducting Order Parameter in $UTe_2$ Determined by Knight Shift Measurement,” *J. Phys. Soc. Jpn.* **91**, 043705 (2022).

[Ghannadzadeh11] S. Ghannadzadeh, M. Coak, I. Franke, P. A. Goddard, J. Singleton, and J. L. Manson, “Measurement of magnetic susceptibility in pulsed magnetic fields using a proximity detector oscillator,” *Rev. Sci. Instrum.* **82**, 113902 (2011).

[Haga22] Y. Haga, P. Opletal, Y. Tokiwa, E. Yamamoto, Y. Tokunaga, S. Kambe, and H. Sakai, “Effect of uranium deficiency on normal and superconducting properties in unconventional superconductor $UTe_2$,” *J. Phys.: Condens. Matter* **34**, 175601 (2022).

[Hamaker79] H. C. Hamaker, L. D. Woolf, H. B. MacKay, Z. Fisk, and M. B. Maple, “Possible observation of the coexistence of superconductivity and long-range magnetic order in $NdRh_4B_4$,” *Solid State Commun.* **31**, 139 (1979).

[Hayes21] I. M. Hayes, D. S. Wei, T. Metz, J. Zhang, Y. S. Eo, S. Ran, S. R. Saha, J. Collini, N. P. Butch, D. F. Agterberg, A. Kapitulnik, and J. Paglione, “Multicomponent superconducting order parameter in $UTe_2$,” *Science* **373**, 797 (2021).

[Helm24] Toni Helm, Motoi Kimata, Kenta Sudo, Atsuhiko Miyata, Julia Stirnat, Tobias Förster, Jacob Hornung, Markus König, Ilya Sheikin, Alexandre Pourret, Gerard Lapertot, Dai Aoki, Georg Knebel, Joachim Wosnitza, and Jean-Pascal Brison, “Field-induced compensation of magnetic exchange as the possible origin of reentrant superconductivity in $UTe_2$,” *Nat. Commun.* **15**, 37 (2024).

[Honda23] Fuminori Honda, Shintaro Kobayashi, Naomi Kawamura, Saori Kawaguchi, Takatsugu Koizumi, Yoshiki J. Sato, Yoshiya Homma, Naoki Ishimatsu, Jun Gouchi, Yoshiya Uwatoko, Hisatomo Harima, Jacques Flouquet, and Dai Aoki, “Pressure-induced structural transition and new superconducting phase in $UTe_2$,” *J. Phys. Soc. Jpn.* **92**, 044702 (2023).

[Huston22] Larissa Q. Huston, Dmitry Y. Popov, Ashley Weiland, Mitchell M. Bordelon, Priscila F. S. Rosa, Richard L. Rowland, II, Brian L. Scott, Guoyin Shen, Changyong Park, Eric K. Moss, S. M. Thomas, J. D. Thompson, Blake T. Sturtevant, and Eric D. Bauer, “Metastable phase of $UTe_2$ formed under high pressure above 5 GPa,” *Phys. Rev. Mat.* **6**, 114801 (2022).

[Jaccarino62] V. Jaccarino and M. Peter, “Ultra-high-field Superconductivity,” *Phys. Rev. Lett.* **9**, 290(1962).

[Jiao20] L. Jiao, S. Howard, S. Ran, Z. Wang, J. O. Rodriguez, M. Sigrist, Z. Wang, N. P. Butch, and V. Madhavan, “Chiral superconductivity in heavy-fermion metal $UTe_2$,” *Nature* **579**, 523 (2020).

[Kallin09] C. Kallin and A. J. Berlinsky, “Is $Sr_2RuO_4$ a chiral p-wave superconductor?” *J. Phys.: Condens. Matter* **21**, 164210 (2009).

[Kinjo23] K. Kinjo, H. Fujibayashi, S. Kitagawa, K. Ishida, Y. Tokunaga, H. Sakai, S. Kambe, A. Nakamura, Y. Shimizu, Y. Homma, D. X. Li, F. Honda, D. Aoki, K. Hiraki, M. Kimata, and T. Sasaki, “Change of superconducting character in $UTe_2$ induced by magnetic field,” *Phys. Rev. B* **107**, L060502 (2023).

[Knafo21a] W. Knafo, M. Nardone, M. Vališka, A. Zitouni, G. Lapertot, D. Aoki, G. Knebel, and D. Braithwaite. “Comparison of two superconducting phases induced by a magnetic field in $UTe_2$,” *Commun. Phys.* **4**, 40 (2021).

[Knafo21b] W. Knafo, G. Knebel, P. Steffens, K. Kaneko, A. Rosuel, J-P. Brison, J. Flouquet, D. Aoki, G. Lapertot, and S. Raymond, “Low-dimensional antiferromagnetic fluctuations in the heavy-fermion paramagnetic ladder $UTe_2$,” *Phys. Rev. B* **104**, L100409 (2021).

[Knafo23] W. Knafo, T. Thebault, P. Manuel, D. D. Khalyavin, F. Orlandi, E. Ressouche, K. Beauvois, G. Lapertot, K. Kaneko, D. Aoki, D. Braithwaite, G. Knebel, and S. Raymond, “Incommensurate antiferromagnetism in $UTe_2$ under pressure,” *arXiv preprint arXiv:2311.05455* (2023).

[Knebel19] Georg Knebel, William Knafo, Alexandre Pourret, Qun Niu, Michal Vališka, Daniel Braithwaite, Gérard Lapertot, Marc Nardone, Abdelaziz Zitouni, Sanu Mishra, Ilya Sheikin, Gabriel Seyfarth, Jean-Pascal Brison, Dai Aoki, and Jacques Flouquet, “Field-reentrant superconductivity close to a metamagnetic transition in the heavy-fermion superconductor $UTe_2$,” *J. Phys. Soc. Jpn.* **88**, 063707 (2019).

[Knebel23] G. Knebel, A. Pourret, S. Rousseau, N. Marquardt, D. Braithwaite, F. Honda, D. Aoki, G. Lapertot, W. Knafo, G. Seyfarth, J-P. Brison, and J. Flouquet, “c axis electrical transport at the metamagnetic transition in the heavy-fermion superconductor $UTe_2$ under pressure,” *Phys. Rev. B* **109**, 155103 (2024).

[Lewin23] Sylvia K. Lewin, Corey E. Frank, Sheng Ran, Johnpierre Paglione, and Nicholas P. Butch, “A review of $UTe_2$ at high magnetic fields,” *Rep. Prog. Phys.* **86**, 114501 (2023).

[Lewin24] Sylvia K. Lewin, Josephine J. Yu, Corey E. Frank, David Graf, Patrick Chen, Sheng Ran, Yun Suk Eo, Johnpierre Paglione, S. Raghu, and Nicholas P. Butch, “Field-angle evolution of the superconducting and magnetic phases of $UTe_2$ around the b axis,” *Phys. Rev. B* **110**, 184520 (2024).

[Mackenzie03] A. P. Mackenzie and Y. Maeno, “The superconductivity of $Sr_2RuO_4$ and the physics of spin triplet pairing,” *Rev. Mod. Phys.* **75**, 657 (2003).

[Maple85] M. Brian Maple, “Induction of superconductivity by applied magnetic fields,” *Nature* **315**, 95 (1985).

[Meul84] H. W. Meul, C. Rossel, M. Decroux, O. Fischer, G. Remenyi, and A. Briggs, “Observation of magnetic-field-induced superconductivity,” *Phys. Rev. Lett.* **53**, 497 (1984).

[Nakamine21] G. Nakamine, K. Kinjo, S. Kitagawa, K. Ishida, Y. Tokunaga, H. Sakai, S. Kambe, A. Nakamura, Y. Shimizu, Y. Homma, D. Li, F. Honda, and D. Aoki,

"Anisotropic response of spin susceptibility in the superconducting state of $UTe_2$ probed with $^{125}$Te-NMR measurement," *Phys. Rev. B* **103**, L100503 (2021).

[Raymond21] S. Raymond, W. Knafo, G. Knebel, K. Kaneko, J.-P. Brison, J. Flouquet, D. Aoki, and G. Lapertot, "Feedback of Superconductivity on the Magnetic Excitation Spectrum of $UTe_2$," *J. Phys. Soc. Jpn.* **90**, 113706 (2021).

[Ran19a] Sheng Ran, Chris Eckberg, Qing-Ping Ding, Yuji Furukawa, Tristin Metz, Shanta R. Saha, I-Lin Liu, Mark Zic, Hyunsoo Kim, Johnpierre Paglione, and Nicholas P. Butch, "Nearly ferromagnetic spin-triplet superconductivity," *Science* **365**, 684-687 (2019).

[Ran19b] Sheng Ran, I-Lin Liu, Yun Suk Eo, Daniel J. Campbell, Paul M. Neves, Wesley T. Fuhrman, Shanta R. Saha, Christopher Eckberg, Hyunsoo Kim, David Graf, Fedor Balakirev, John Singleton, Johnpierre Paglione, and Nicholas P. Butch, "Extreme magnetic field-boosted superconductivity," *Nat. Phys.* **15**, 1250 (2019).

[Ran21a] Sheng Ran, Shanta R. Saha, I-Lin Liu, David Graf, Johnpierre Paglione, and Nicholas P. Butch, "Expansion of the high field-boosted superconductivity in $UTe_2$ under pressure," *npj Quant. Matls.* **6**, 75 (2021).

[Ran21b] Sheng Ran, I-Lin Liu, Shanta R. Saha, Prathum Saraf, Johnpierre Paglione, and Nicholas P. Butch, "Comparison of two different synthesis methods of single crystals of superconducting uranium ditelluride," *J. Vis. Exp.***173**, e62563 (2021).

[Raymond21] Stephane Raymond, William Knafo, Georg Knebel, Koji Kaneko, Jean-Pascal Brison, Jacques Flouquet, Dai Aoki, and Gerard Lapertot, "Feedback of superconductivity on the magnetic excitation spectrum of $UTe_2$" *J. Phys. Soc Jpn.* **90**, 113706 (2021).

[Rosa22] P. F. S. Rosa, A. Weiland, S. S. Fender, B. L. Scott, F. Ronning, J. D. Thompson, E. D. Bauer, and S. M. Thomas, "Single thermodynamic transition at 2 K in superconducting $UTe_2$ single crystals," *Commun. Mater.* **3**, 33 (2022).

[Rosuel23] A. Rosuel, C. Marcenat, G. Knebel, T. Klein, A. Pourret, N. Marquardt, Q. Niu, S. Rousseau, A. Demuer, G. Seyfarth, G. Lapertot, D. Aoki, D. Braithwaite, J. Flouquet, and J. P. Brison, "Field-induced tuning of the pairing state in a superconductor," *Phys. Rev. X* **13**, 011022 (2023).

[Sakai22] H. Sakai, P. Opletal, Y. Tokiwa, E. Yamamoto, Y. Tokunaga, S. Kambe, and Y. Haga, "Single crystal growth of superconducting $UTe_2$ by molten salt flux method," *Phys. Rev. Mater.* **6**, 073401 (2022).

[Sakai23] H. Sakai, Y. Tokiwa, P. Opletal, M. Kimata, S. Awaji, T. Sasaki, D. Aoki, S. Kambe, Y. Tokunaga, and Y. Haga, "Field-induced multiple superconducting phases in $UTe_2$ along hard magnetic axis," *Phys. Rev. Lett.* **130**, 196002 (2023).

[Sau12] J. D. Sau and S. Tewari, “Topologically protected surface Majorana arcs and bulk Weyl fermions in ferromagnetic superconductors,” *Phys. Rev. B* **86**, 104509 (2012).

[Shishidou21] Tatsuya Shishidou, Han Gyeol Suh, P. M. R. Brydon, Michael Weinert, and Daniel F. Agterberg, “Topological band and superconductivity in $UTe_2$,” *Phys. Rev. B* **103**, 104504 (2021).

[Singleton00], J. Singleton, J. A. Symington, M-S. Nam, A. Ardavan, M. Kurmoo, and P. Day, “Observation of the Fulde–Ferrell–Larkin–Ovchinnikov state in the quasi-two dimensional organic superconductor κ–$(BEDT–TTF)_2Cu(NCS)_2$,” *J. Phys.: Condens. Matter* **12**, L641 (2000).

[Thomas20] S. M. Thomas, F. B. Santos, M. H. Christensen, T. Asaba, F. Ronning, J. D. Thompson, E. D. Bauer, R. M. Fernandes, G. Fabbris, and P. F. S. Rosa, “Evidence for a pressure-induced antiferromagnetic quantum critical point in intermediate-valence $UTe_2$,” *Sci. Adv.* **6**, eabc8709 (2020).

[Thomas21] Sean Michael Thomas, Callum Stevens, Frederico B. Santos, Shannon Sanaya Fender, Eric Dietzgen Bauer, Filip Ronning, Joe David Thompson, Andrew Huxley, and P. F. S. Rosa, “Spatially inhomogeneous superconductivity in $UTe_2$,” *Phys. Rev. B* **104**, 224501 (2021).

[Uji01] S. Uji, H. Shinagawa, T. Terashima, T. Yakabe, Y. Terai, M. Tokumoto, A. Kobayashi, H. Tanaka, and H. Kobayashi, “Magnetic-field-induced superconductivity in a two-dimensional organic conductor,” *Nature* (London) **410**, 908 (2001).

[Wei22] D. S. Wei, D. Saykin, O. Y. Miller, S. Ran, S. R. Saha, D. F. Agterberg, J. Schmalian, N. P. Butch, J. Paglione, and A. Kapitulnik, “Interplay between magnetism and superconductivity in $UTe_2$,” *Phys. Rev. B* **105**, 024521 (2022).

[Wilhelm23] F. Wilhelm, J. P. Sanchez, D. Braithwaite, G. Knebel, G. Lapertot, and A. Rogalev, “Investigating the electronic states of $UTe_2$ using X-ray spectroscopy,” *Commun. Phys.* **6**, 1 (2023).

[Wolowiec15] C. T. Wolowiec, B. D. White, and M. B. Maple, “Conventional magnetic superconductors,” *Physica C* **514**, 113 (2015).

[Wu24] Z. Wu, T. I. Weinberger, J. Chen, A. Cabala, D. V. Chichinadze, D. Shaffer, J. Pospíšil, J. Prokleška, T. Haidamak, G. Bastien, V. Sechovský, A. J. Hickey, M. J. Mancera-Ugarte, S. Benjamin, D. E. Graf, Y. Skourski, G. G. Lonzarich, M. Vališka, F. M. Grosche, and A. G. Eaton, “Enhanced triplet superconductivity in next-generation ultraclean $UTe_2$,” *Proc. Nat. Acad. Sci.* **121**, e2403067121 (2024).

**Figures (**High-Magnetic Field Phases in $U_{1-x}Th_xTe_2$, C. M. Moir *et al.*)

FIGURE 1

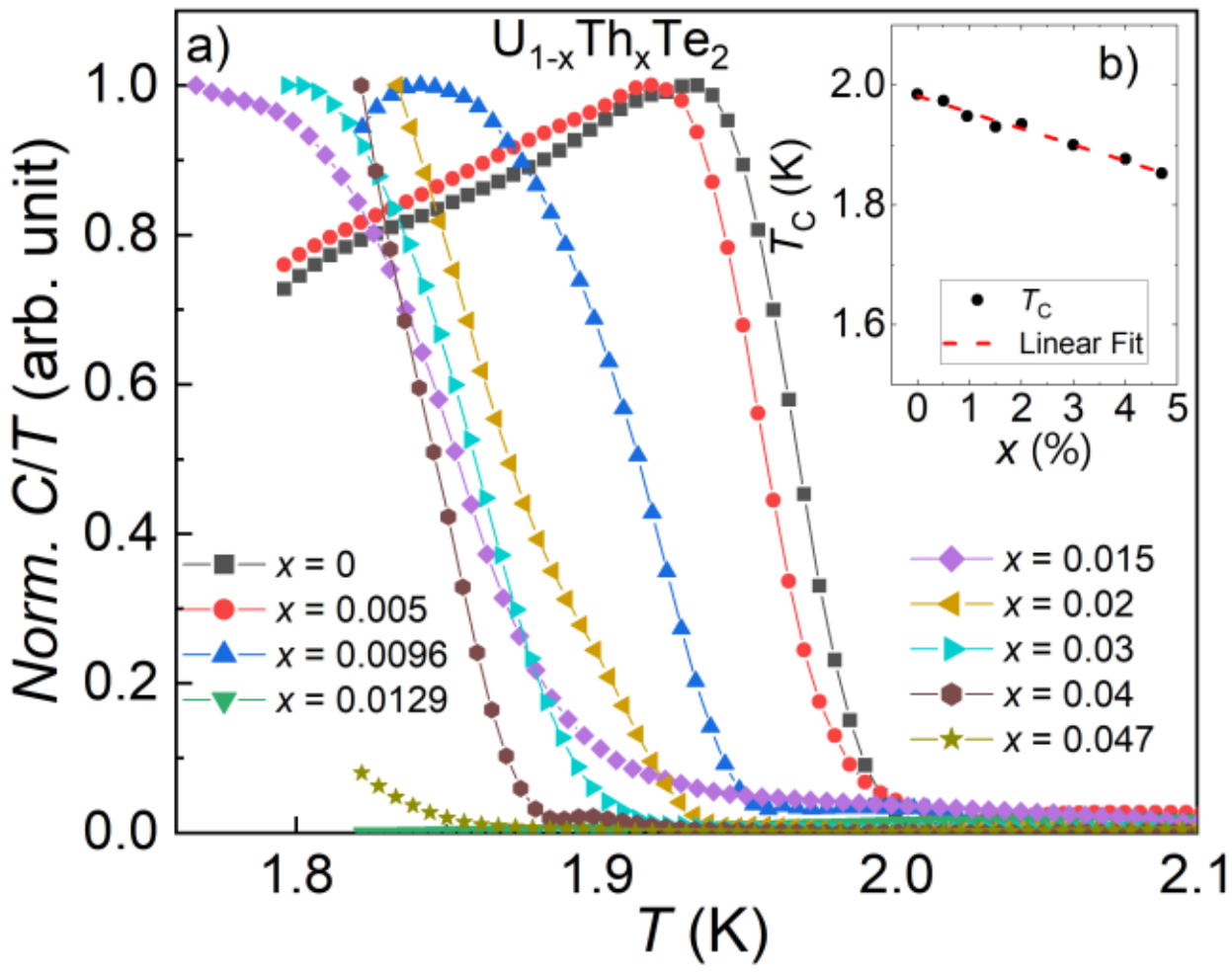

Figure 1: (a) Normalized *C*/*T* as a function of temperature (*T*) for $U_{1-x}Th_xTe_2$ single crystals with various nominal Th concentrations *x*. Each curve represents a different Th concentration, with *x* values ranging from 0 to 0.047, as indicated in the legend. The inset (b) displays the superconducting transition temperature ($T_c$) as a function of the Th concentration *x*; $T_c$ is defined as the intercept of a straight line that is tangent to the *C*/*T* curve in the region of the superconducting "jump" in C/*T* on the *T*-axis, indicating the onset of the superconducting transition. The value of $T_c$ decreases linearly with Th concentration at a rate $dT_c/dx \approx -0.03$ K/% Th. No anomaly is visible for the sample with $x = 0.0129$ because this sample was grown using a higher temperature gradient in the CVT process which produces crystals that have $T_c$'s below 1.8 K, even for the parent compound. [Rosa22]

FIGURE 2

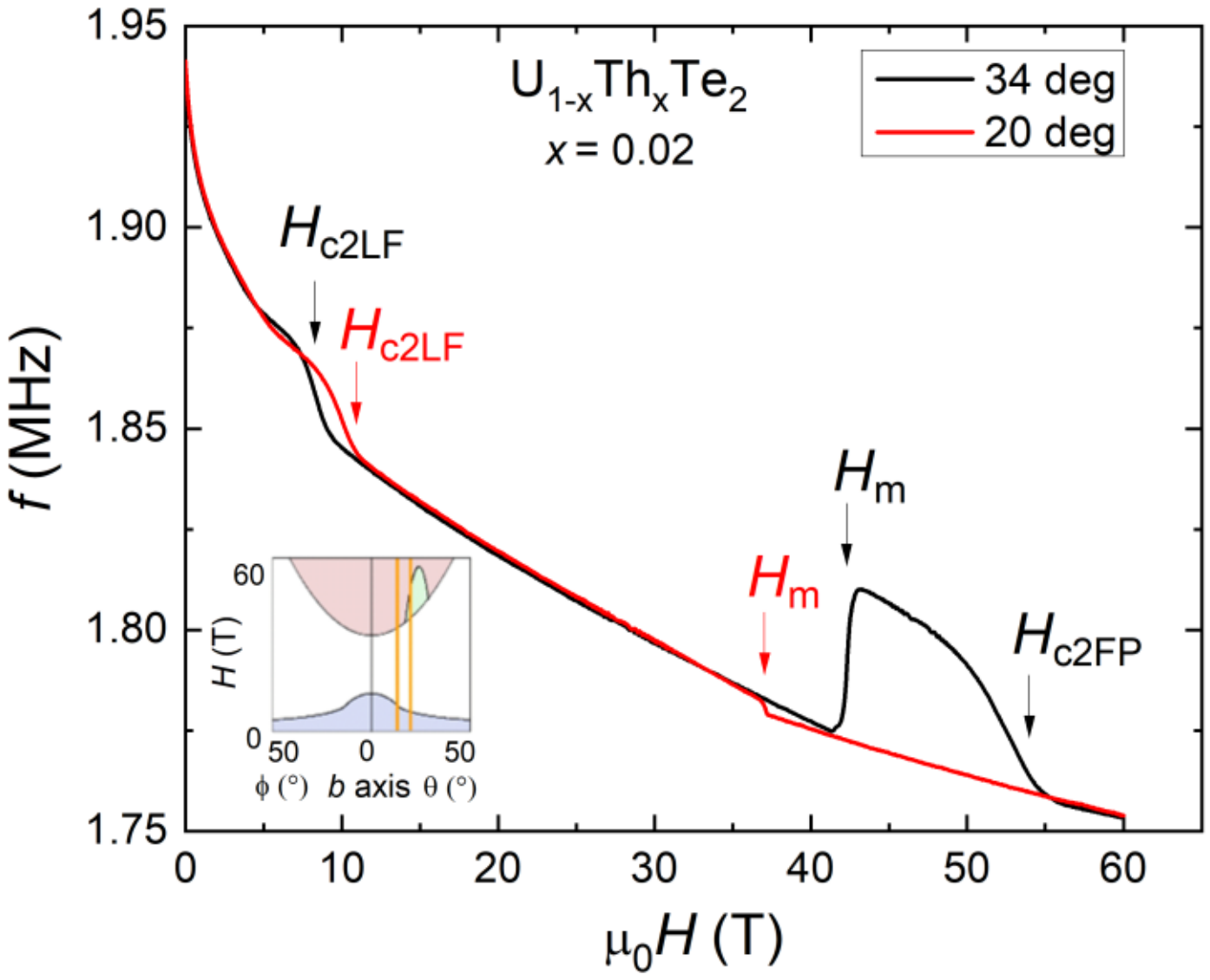

Figure 2: PDO frequency as a function of magnetic field for a $U_{1-x}Th_xTe_2$ single crystal with $x = 0.02$ at angles $\theta$ of 20° (red) and 34° (black) from the *b*-axis in the (*b*, *c*) plane. The changes in frequency are indicated with arrows for identification. At both angles, the critical field for the low field superconducting phase, $H_{c2LF}$, is close to 10 T. At 34° from the *b*-axis, we see the onset of the $SC_{FP}$ and FP phase, $H_m$, at roughly 40 T and the suppression of the same phase at nearly 55 T, $H_{c2FP}$. At 20°, we do not see evidence of the $SC_{FP}$ phase, while at 34°, we do observe the onset of the field polarized FP phase, which is obscured by the $SC_{FP}$ phase at 34°.

FIGURE 3

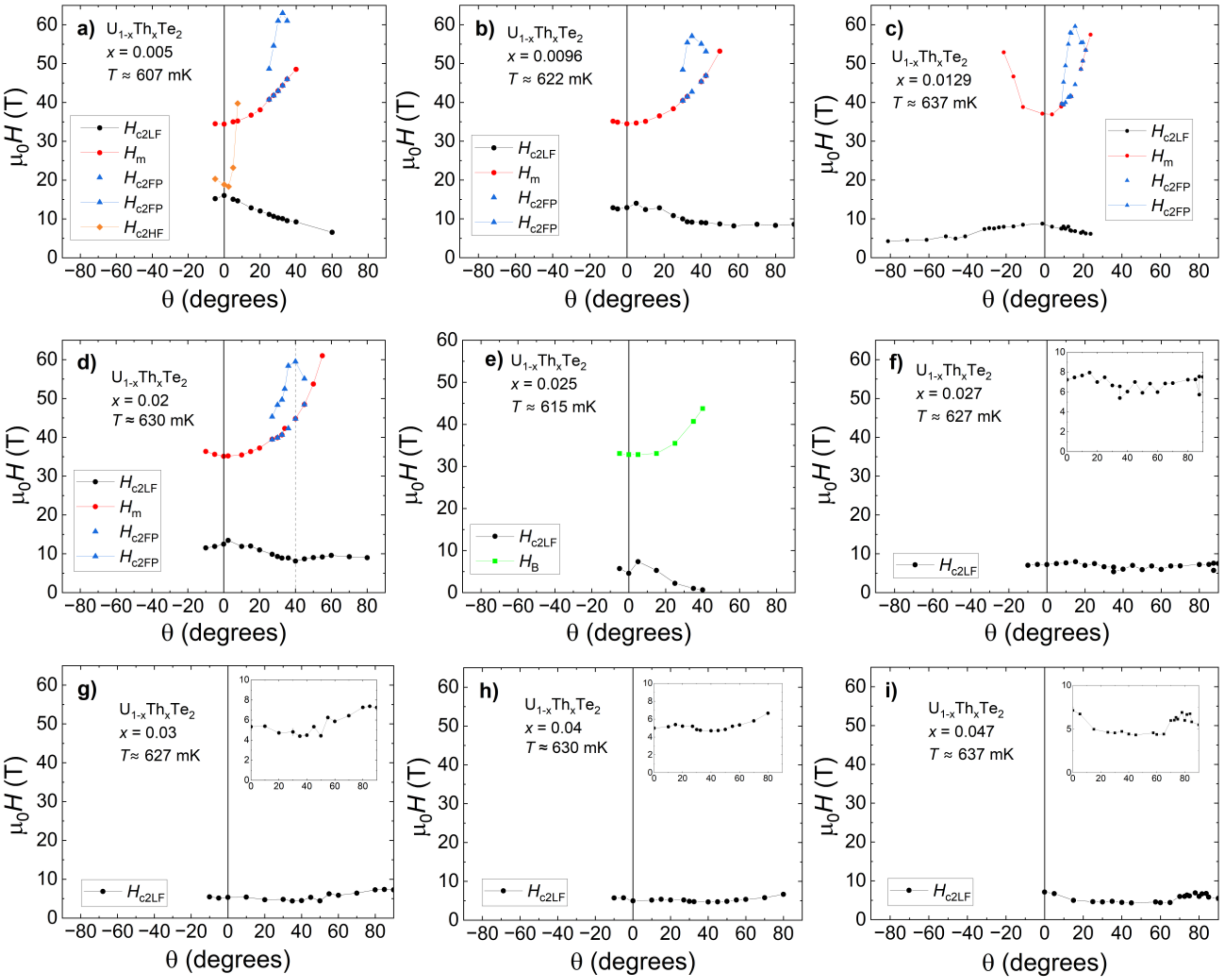

Figure 3: Phase diagrams of magnetic field vs. angle $\theta$ from the *b*-axis in the (*b*, *c*) plane (angles from (010) to (001)) at base temperature of ~0.6 K for $U_{1-x}Th_xTe_2$ single crystals with $x$ = 0.005, 0.0096, 0.0129, 0.02, 0.025, 0.027, 0.03 0.04, and 0.047. Panels a-d): All superconducting phases seen at $x = 0$ are present except for the $SC_{HF}$ phase that occurs in the vicinity of the *b*-axis. In a) there is a feature close to the *b*-axis that may be a remnant of the $SC_{HF}$ phase. c) The FP and $SC_{FP}$ phases for $x = 0.0129$ appear compressed in the x-direction compared to those for $x = 0$ and 0.02, likely due to a small projection towards the *a*-axis. We note that for $x = 0.02$, there is a minimum in the $SC_{LF}$ phase upper critical field $H_{c2LF}$ that occurs at the same angle where $H_{c2FP}$ of the $SC_{FP}$ phase is at a maximum. e) We no longer observe the features associated with the FP or $SC_{FP}$ phases. Instead, we notice that there is a broad feature in the data that occurs at similar fields and angles as $H_m$. f-i) Only the $SC_{LF}$ phase is observed for $x > 0.03$. For both samples, $H_{c2HF}$ and $H_{c2LF}$ are maximal at the *b*- and *c*-axes, respectively.

FIGURE 4

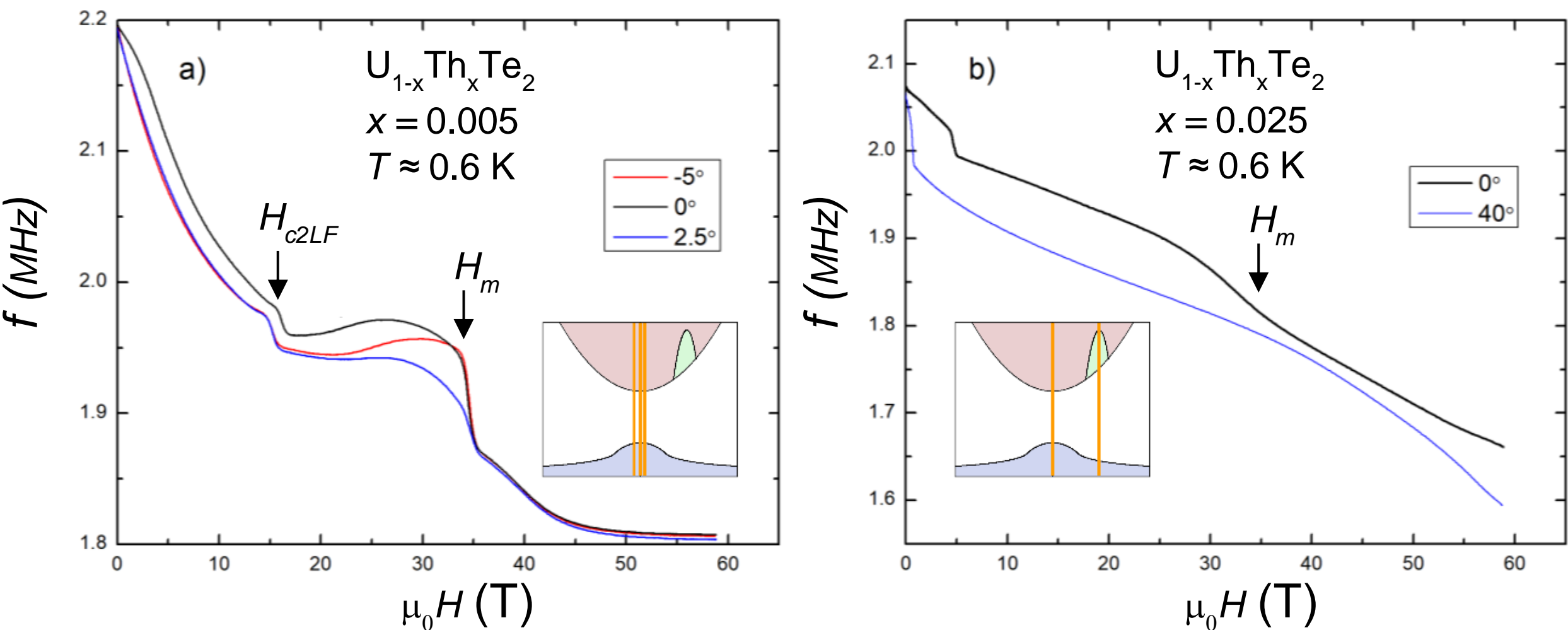

Figure 4: PDO frequency as a function of magnetic field demonstrating different behaviors specific to $U_{1-x}Th_xTe_2$ single crystals with $x = 0.005$ and 0.025. a) Data taken close to the *b*-axis of $x = 0.005$ showing an increase in the frequency between $H_{c2HF}$ and $H_m$, possibly indicating a trace of re-entrant superconductivity. b) Data trace providing an example of the broad feature at the *b*-axis and a curve showing the absence of the feature below 60 T at 40° from the *b*-axis to the *c*-axis. Insets show the measurement angles of data presented on a cartoon phase diagram of the phases present below $x = 0.03$.

FIGURE 5

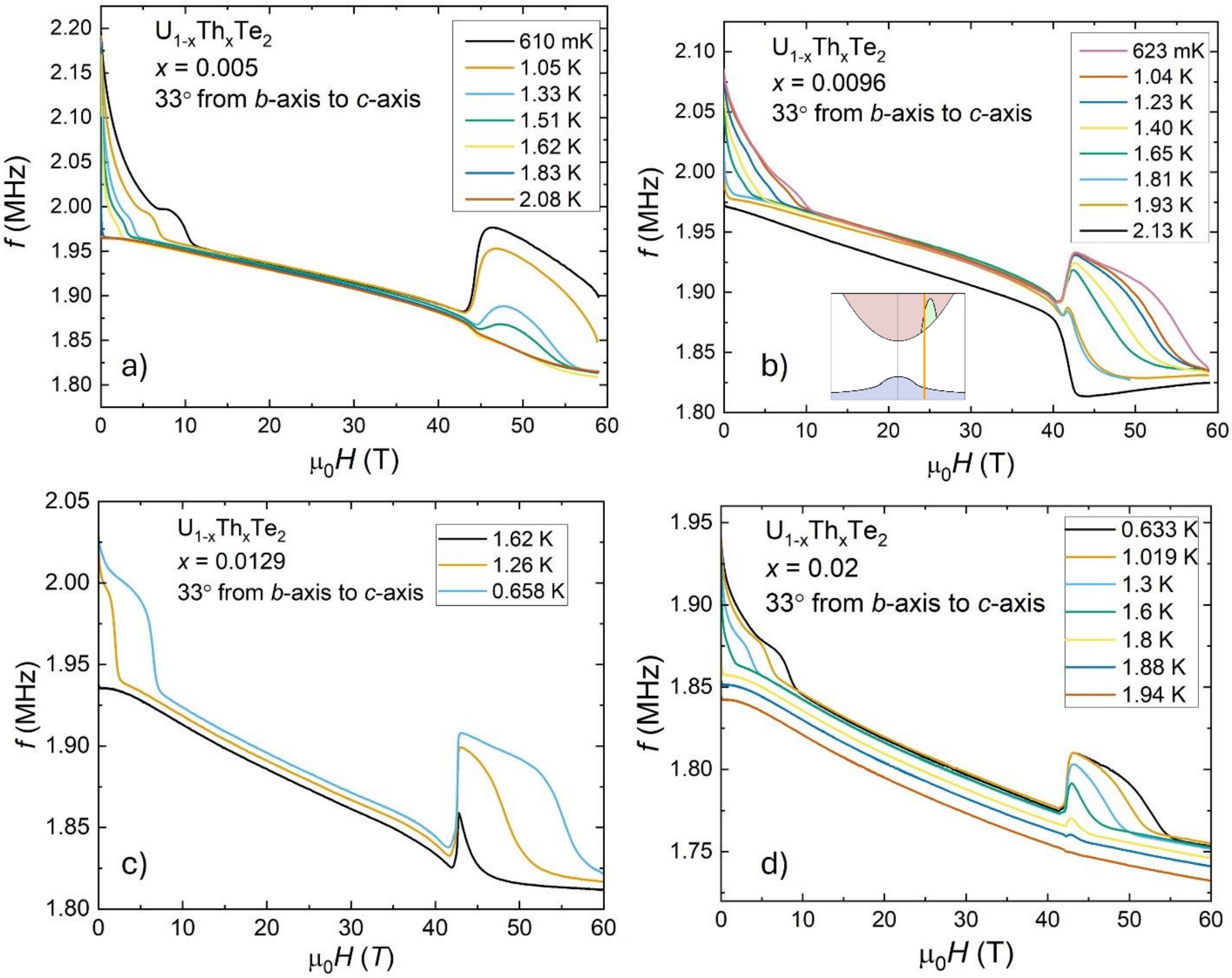

Fig 5: PDO frequency as a function of applied magnetic field for $U_{1-x}Th_xTe_2$ single crystals with *x* = 0.005, 0.0096, 0.0129, and 0.02 at several temperatures and a fixed angle from the *b*-axis to the *c*-axis. Angles were chosen such that the $SC_{FP}$ phase would be intersected. For all samples, the $SC_{FP}$ phase is still observed after the zero-field SC phase has been completely suppressed. For samples where temperatures were high enough to suppress the $SC_{FP}$ phase, the meta-magnetic transition into the FP phase becomes visible at the onset field for the $SC_{FP}$ phase. The meta-magnetic transition appears to be temperature independent up to roughly 2 K. We note for *x* = 0.005, a small feature similar to a superconducting phase appears before $H_{c2LF}$, which may be a remnant of the re-entrant superconducting phase that occurs along the *b*-axis in the parent compound; however, further investigations are needed. The inset in b) shows the angle at which measurements were made relative to the phases present below *x* = 0.03 for all panels.

FIGURE 6

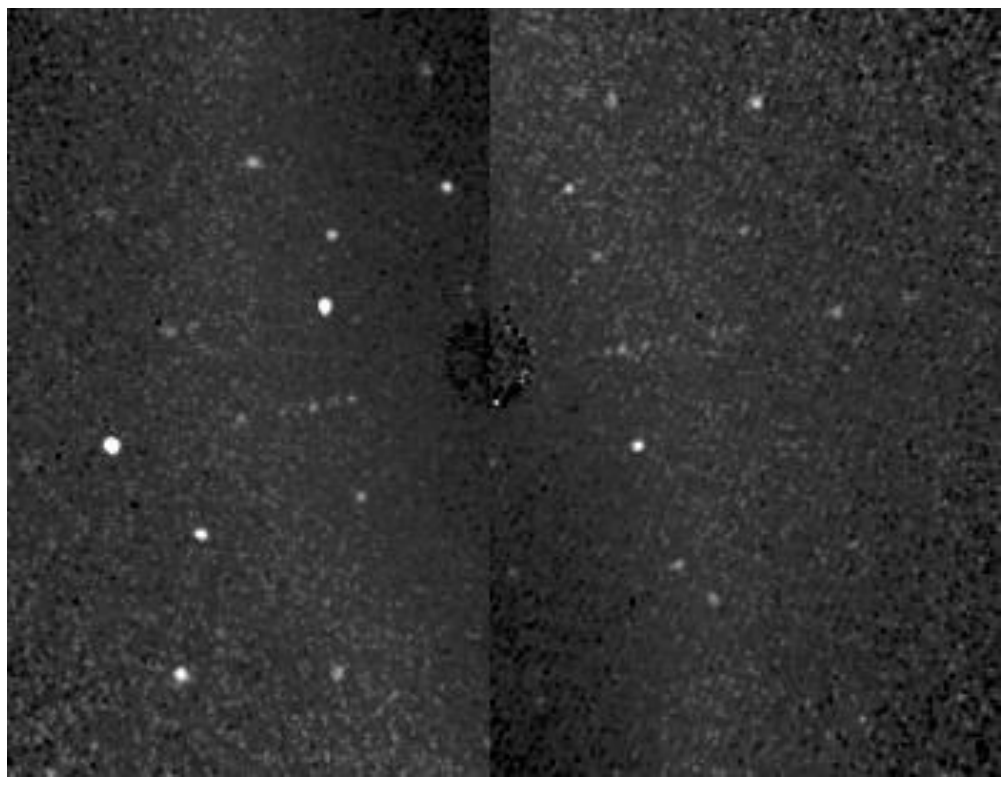

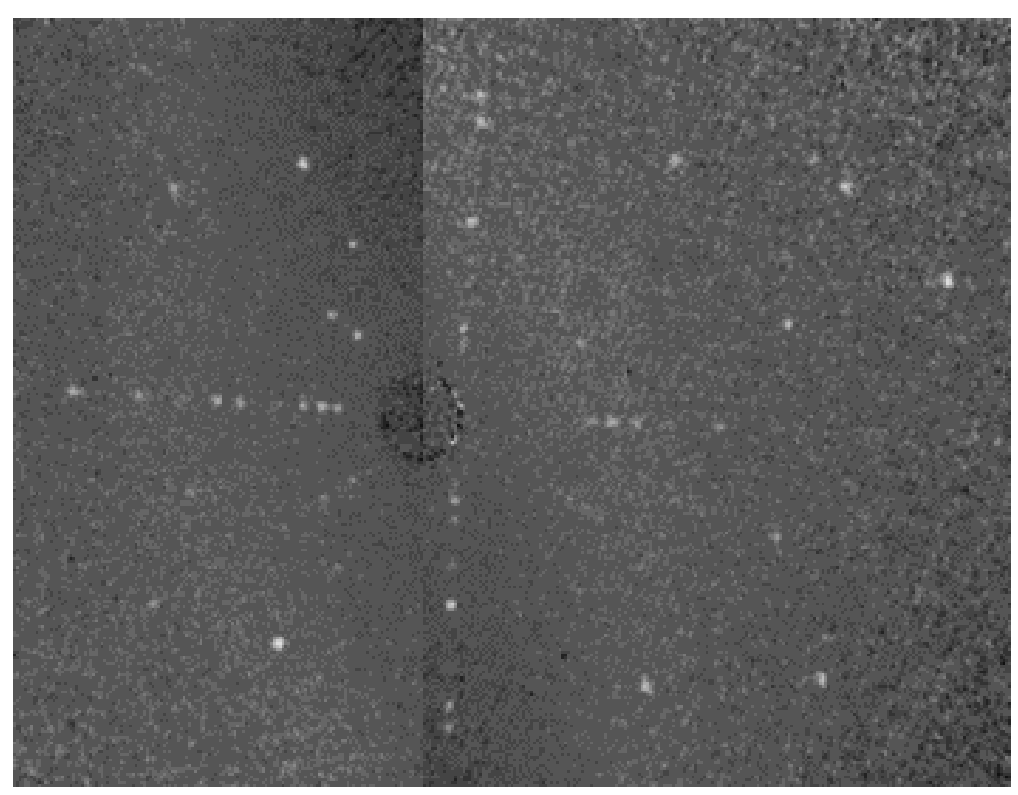

Figure 6: Laue images of two $U_{1-x}Th_xTe_2$ crystals, a) $x = 0$ and b) $x = 0.04$, oriented with the *c*-axis out of the plane. Both patterns show clear features of orthorhombic crystal patterns with no major changes in the patterns, indicating the crystal structure undergoes no large changes from $x = 0$ to $x = 0.04$, near the upper limit of the substitution range in this work.